\documentclass[11pt]{article}
\usepackage{moriond,epsfig}

\def\Journal#1#2#3#4{{#1} {\bf #2}, #3 (#4)}

\def\AAA{\em Astron. \& Astrophys.}
\def\AJ{\em Astron. J.}
\def\APJ{\em Astrophys. J.}
\def\APJS{\em Astrophys. J. Supp.}
\def\MNRAS{\em Mon. Not. Roy. Astron. Soc.}
\def\ARAA{\em Ann. Rev. Astron. Astrophys.}

\def\be{\begin{equation}}
\def\ee{\end{equation}}
\def\bea{\begin{eqnarray}}
\def\eea{\end{eqnarray}}

\newcommand{\Msol}{M_{\odot}}
\newcommand{\inu}{i_\nu}
\newcommand{\jnu}{j_\nu}
\newcommand{\nel}{n_{\rm e}}
\newcommand{\sigT}{\sigma_{\rm T}}
\newcommand{\Snu}{S_\nu}
\newcommand{\Mgas}{M_{\rm gas}}
\newcommand{\Da}{D_{\rm ang}}
\newcommand{\sigY}{\sigma_{\rm Y}}
\newcommand{\thetac}{\theta_{\rm c}}

\begin{document}
\vspace*{4cm}
\title{SZ Surveys are Coming: What should we do?}

\author{ J.G. Bartlett }

\address{APC, 11 pl. Marcelin Berthelot, 75231 Paris Cedex 05, FRANCE (UMR 7164 -- 
	CNRS, Universit\'e Paris 7, CEA, Observatoire de Paris) }

\maketitle\abstracts{
Galaxy clusters furnish extremely rich information on the contents and structure of our universe.  The potential of galaxy cluster studies to constrain dark energy, for example, motivates a number of ambitious cluster surveys.  Among these, surveys based on the Sunyaev-Zel'dovich (SZ) effect are particularly powerful for their ability to cleanly select clusters out to redshifts $z>1$.  Now poised to begin surveying substantial
areas of sky, dedicated interferometers, bolometer cameras and the Planck satellite will soon produce 
large cluster catalogs that will provide a precise measure of the cosmic expansion rate over a range of redshifts and
precipitate a new understanding of structure and galaxy formation.  I review the science potential of
these surveys and examine some issues of SZ cluster catalog construction.
}

\section{Introduction}
Detailed observations of the cosmic microwave background 
anisotropies \cite{cmb_gen}, distant SNIa \cite{snIa_gen} and of the galaxy distribution \cite{lss_gen} 
have driven the tremendous advance in recent years leading to the 
development of a standard cosmological model.  
Ongoing research aims to test the model's coherence and to answer
outstanding fundamental questions: What is the nature of dark matter and of the mysterious dark energy accelerating the present expansion of the universe? What is
the physics of the early universe, in particular of the inflation epoch? How can
we use detailed cosmological observations to probe fundamental physics, such as 
the neutrino sector? And how do galaxies form and evolve in the cosmic web of 
large--scale structure.  These exciting questions inspire large observational 
programs centered on the CMB, SNIa searches and wide--field multiband surveys.  

\begin{figure}
\includegraphics[scale=0.45]{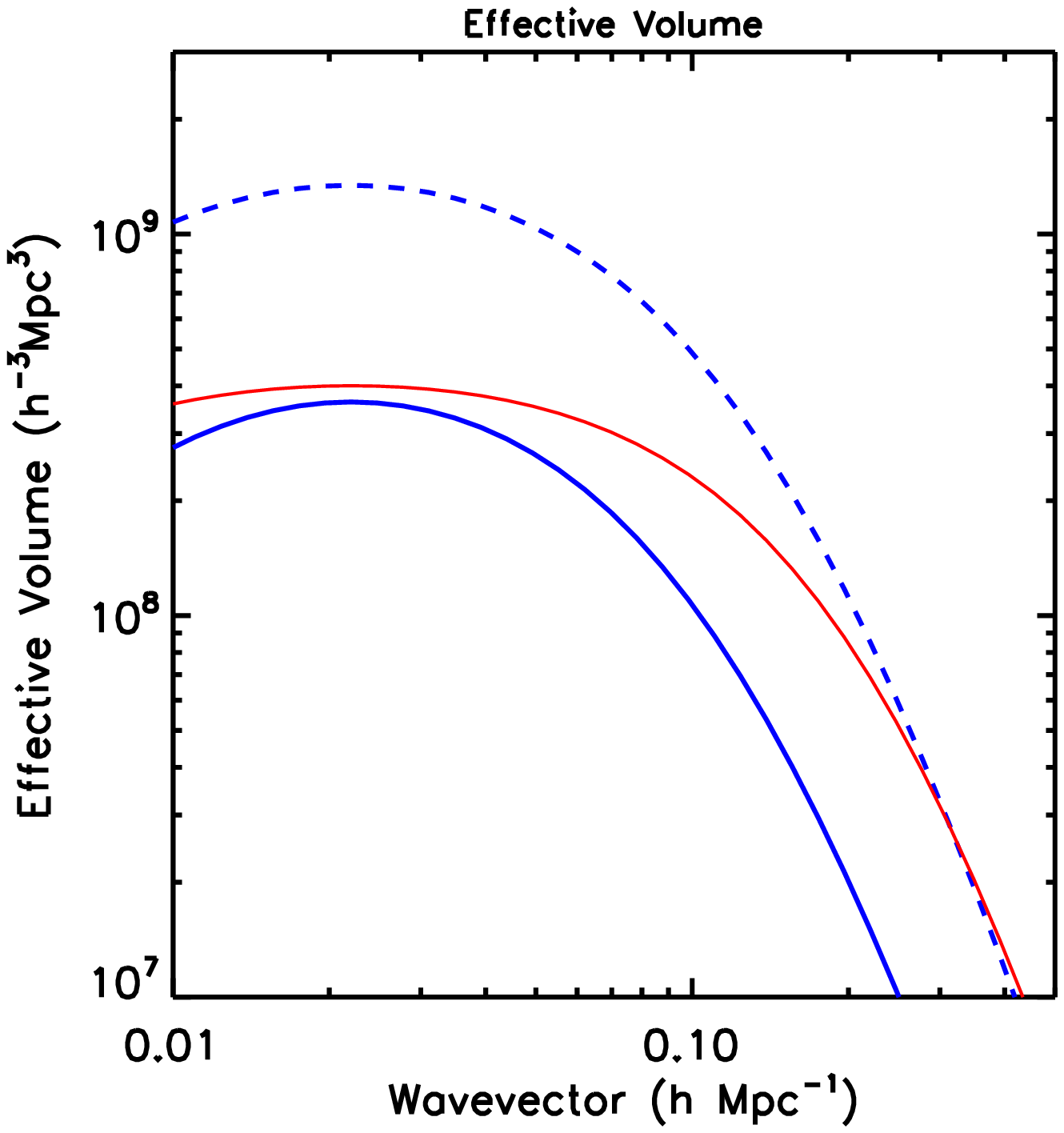}
\includegraphics[scale=0.45]{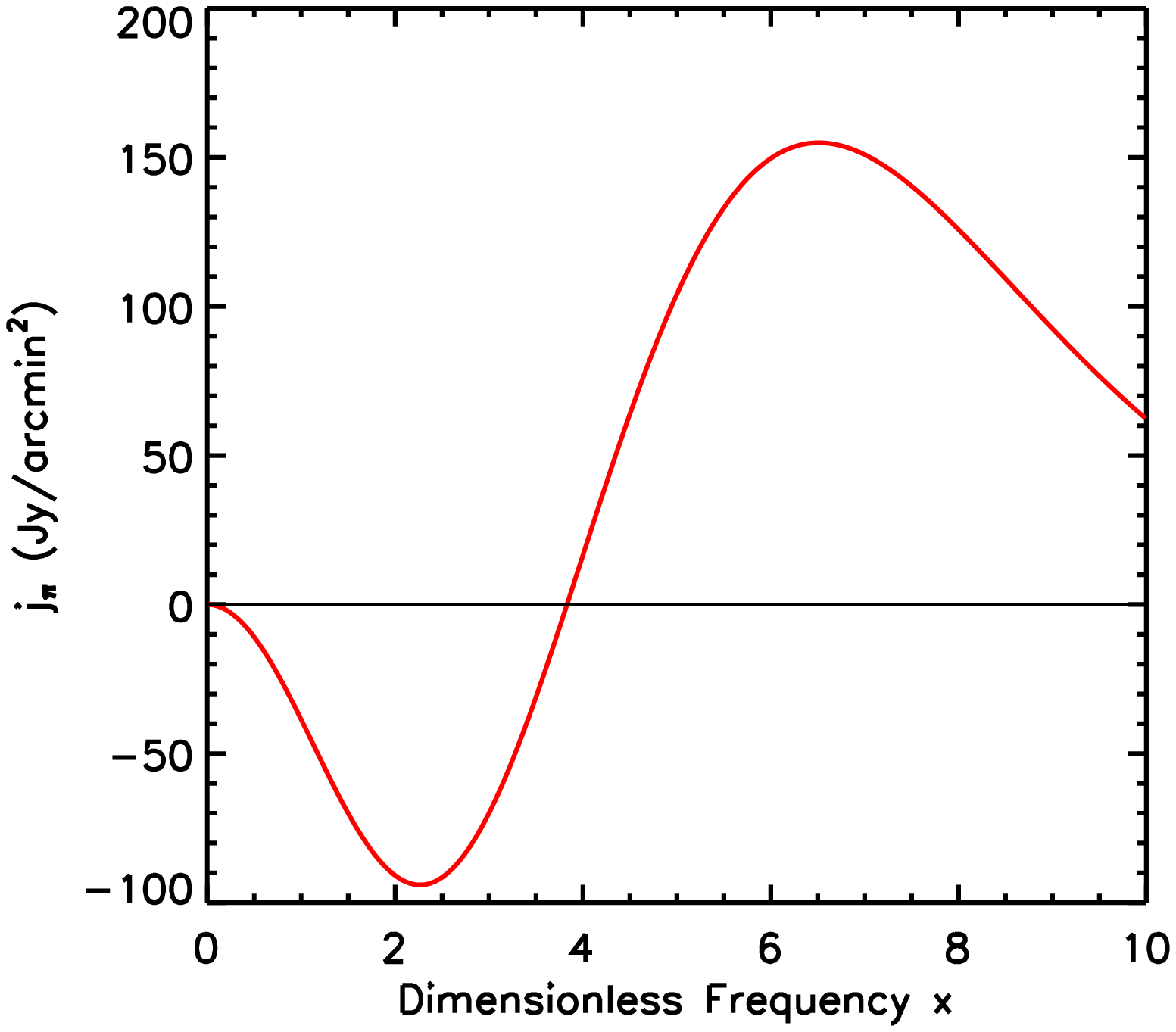}
\caption{{\em Left:} The effective volume as a function of wavenumber for the SDSS LRG sample (middle, red solid curve)
and for two mass--limited cluster samples, one at $M>10^{14}\Msol$ (upper, blue dashed line) and one
at $M>2\times 10^{14}\Msol$ (lower, blue solid line).  In all cases the survey area is 
$\Omega_{\rm survey}=4000$~deg$^2$, corresponding to the actual survey area of the SDSS LRG sample that 
detected the baryon acoustic peak and to the proposed South Pole Telescope SZ survey, which targets these cluster mass limits.
{\em Right:} The thermal SZ spectrum $\jnu$ as a function of dimensionless frequency $x=h\nu/kT_{\rm cmb}$ 
($T_{\rm cmb}=2.725$~K).  The zero crossing is at 217~GHz.
}
\label{fig:Veff}
\end{figure}

Galaxy clusters offer a unique avenue of attack on several of these key
questions:
\begin{enumerate}
\item Their abundance and evolution with redshift are 
highly sensitive to the statistics and growth rate of the density perturbations, which in turn depends on the cosmological parameters;
\item They are ideal tracers of the largest scale structures.  With a mean 
separation of $\sim 50$~Mpc, they efficiently sample structures of wavelength
$\sim 100$~Mpc and larger, such as the baryon acoustic peaks in the matter 
power spectrum (see Figure~\ref{fig:Veff}) \cite{bao};
\item They provide a well--defined, quasi--closed environment for galaxy formation
studies.  In the cluster environment we directly observe the stellar, 
diffuse gas and dark matter components of the cosmic fluid;  
\item Combined X--ray and millimeter cluster observations (the Sunyaev-Zel'dovich 
effect; see below) permit distance measurements and hence the construction of a Hubble
diagram with which to measure the matter density and dark energy abundance and 
equation--of--state (like SNIa distance measurements).
\end{enumerate}

\section{Cluster Surveys}

Tapping the cosmological potential of galaxy clusters requires large, homogeneous
cluster catalogs extending to redshifts greater than unity.  The strongest
constraints on dark energy, for example, will come from observed cluster evolution 
in the transition between matter and dark energy domination around $z=1$ and beyond.  
Note that today there are only of order 10 clusters
with spectroscopically confirmed redshifts $z>1$~\cite{clusters_at_high_z}.  
The need for deeper and larger
catalogs motivates a number of substantial observational efforts 
in different wavebands.  

We may search for clusters in a variety of ways:
\begin{enumerate}
\item As overdensities in the galaxy population.  This may either be done in 
redshift space, or in color space using multiband imaging which can 
provide estimated photometric redshifts over large areas of sky; 
\item By their gravitational lensing of background galaxy images.  With a large imaging
survey, we could search for clusters both via their lensing signal and in color space (as above);
\item By the X--ray emission from their intracluster medium (ICM), which at keV temperatures 
emits copiously in X-rays.  Cluster surveys and observations in the X--ray band with
satellites like ROSAT, Chandra and XMM have significantly advanced our 
understanding of clusters and their evolution, and have played an important
role in establishing the standard cosmological model~\cite{rosati_rev}.
\item By the Sunyaev--Zel'dovich (SZ) effect, a distortion of the
CMB frequency spectrum caused by a transfer of energy from the hot ICM electrons
to CMB photons via Compton scattering.
\end{enumerate}

Each of these techniques has its own advantages and disadvantages.  
To appreciate them we should situate clusters in the standard 
theoretical framework that describes them as single, massive dark matter 
halos containing hot gas and galaxies with their subhalos.  
Dominated by dark matter and its 
gravitational evolution, clusters are fundamentally described by their halo mass 
and redshift.  When building a cluster catalog, therefore, one should quantify
its properties in terms of these basic parameters in order for proper comparison
to theory.  Specifically, we must accurately calculate the survey selection
function and determine the observable--mass relation.  Doing the latter is 
often referred to as calibrating the observable--mass relation, or for short, 
survey calibration.

Large area redshift surveys are prohibitively expensive.  Less expensive, imaging surveys
rely on a consistent relation between the parent cluster halo and its 
member galaxy population; in particular, the relation between cluster mass and 
redshift, and galaxy colors and luminosity.  Therein lies a potentially large 
uncertainty due to our limited understanding of galaxy formation and evolution 
(which these cluster surveys are designed to help us improve!).

Lensing--based cluster surveys are appealing because they select clusters on 
their projected mass density.  They will, however, suffer from the projection
of mass along the line--of--sight, and their detection efficiency depends on the
redshift distribution of the lensed galaxies.  The latter is relatively easily 
controlled with photometric redshift information, while the impact of projection effects
on survey selection functions and mass recovery requires further study~\cite{lensing_proj}.  

The properties of the hot ICM are closely linked to those of the cluster gravitational
potential, implying that X--ray surveys select clusters based on their
dark matter halos, similarly to lensing surveys.  In addition, X--ray surveys 
do not suffer from projection effects, because cluster X--ray emission varies as the square of the gas density.  For this same reason, however, the cluster X--ray emission
is dominated by the core regions and hence physics that is difficult to model.
Furthermore, X--ray surveys require large, expensive satellite missions.
Not designed as survey instruments, Chandra and XMM--Newton will provide only
relatively limited area surveys~\cite{xmm_seren}.  

Our focus here is on the fourth method based on the SZ effect, which possesses 
a number of advantageous properties for cluster surveying, as I detail below.
As for X--ray observations, the SZ effect selects clusters based on
their ICM properties and potential well; and like lensing surveys, SZ surveys
tend to suffer from projection effects, since the SZ signal is proportional to 
the projected gas mass density.  
A number of planned SZ surveys have heightened 
anticipation as they prepare to observe large areas of sky over the next 
few years.  These include ground--based instruments starting operations this year, as well
as the Planck satellite scheduled for launch in 2008 and which will cover the 
entire sky.  

\section{The Sunyaev-Zel'dovich Effect}
The SZ signal arises from the energy transfer via Compton scattering
from hot electrons in the ICM to CMB photons traversing the cluster~\cite{sz}.
Conserving photon number, the scattering diffuses photons from low to high
energies and creates a unique frequency--dependent spectral distortion that is negative
at low frequencies and positive at high frequencies.
We quantify the distortion by the change in observed sky brightness towards
the cluster relative to the unperturbed CMB seen from surrounding blank sky:
\begin{equation}
\Delta \inu = y \jnu
\end{equation}
where the Compton--$y$ parameter specifies the amplitude of the 
distortion as an integral of the ICM pressure along the line--of-sight:
\begin{equation}
y \equiv \int d\tau\; \frac{kT}{mc^2} = \int dl\; \frac{kT}{mc^2}\nel\sigT
\end{equation}
with $T$ being the electron (ICM) temperature, $m$ the electron mass, $\nel$
the electron number density and $\sigT$ the Thomson scattering cross section
(we're in the non--relativistic limit of Compton scattering).
The function $\jnu$ gives the frequency dependence of the spectral distortion and
is universal to all clusters at all redshifts (See Figure~\ref{fig:Veff}).  It is negative below 217~GHz, where
the effect is a brightness decrement relative to the mean sky brightness (the CMB),
and positive above, where the effect produces an excess brightness.  

We refer to the total {\em SZ flux} as 
\begin{equation}
\Snu = Y\jnu
\end{equation}
where the integrated Compton parameter is 
\begin{equation}
Y = \int d\Omega\; y \propto \frac{\Mgas T}{\Da^2(z)}
\end{equation}

\section{SZ Cluster Surveys}
The SZ signal has a number of virtues when considered as a means of cluster surveying.
Firstly, the surface brightness $\Delta\inu$ is independent of distance and depends only on 
intrinsic cluster properties; this is also manifest by the fact that $Y$ varies with the
{\em angular diameter} distance, as opposed to the luminosity distance.  Secondly, the spectral
signature is unique, unlike any other astrophysical source, and it is universal to all clusters.

Thirdly, the total SZ flux $\Snu$ is directly proportional to the total thermal energy of 
the ICM, a quantity fixed by energetics during cluster formation and independent of the final 
spatial and temperature structure of the gas.  For this reason, we expect the SZ signal to
tightly correlate to cluster virial mass, as indeed borne out in numerical simulations;  for example, Motl et al.~\cite{motl_etal05} 
simulated examined different simulations, including a variety of cooling and feedback mechanisms, and find 
only a $\sim 5$\% scatter in $Y$ at a given mass.

These characteristics of the SZ signal lead to one to expect that a SZ survey will select clusters based on their
mass, and that the limiting mass will be nearly constant with redshift.  This is a valuable property for 
studying any evolutionary trends, for we can compare similar mass objects at different epochs.

\begin{table}[htbp]
\centering
\begin{tabular}{|c|c|c|c|c|c|c|}\hline
  Name & Frequencies & Res. {\it fwhm} & Start & Inst. noise          & Surf. dens. (5$\sigma$)  & Survey Area        \\
    \  &    [GHz]    &   [arcmin]      & date  & [$\mu {\rm K/beam}$] & [deg$^{-2}$]             & [${\rm deg}^{2}$]  \\ \hline 
\multicolumn{7}{|c|}{Interferometers}\\ \hline
   AMI &      15     &  1--2           &  2006 &         8            & 16                       &      10            \\ \hline
 AMiBA &      95     & $\sim 2$        &  2006 & \multicolumn{3}{c|}{}\\ \hline
  SZA &    30 (+90)  & $\sim 1$        &  2006 & \multicolumn{3}{c|}{}\\ \hline
\multicolumn{7}{|c|}{Bolometers}\\ \hline
   ACT &     145     &       1.7       & 2006  &      1.7             & 40                       &     200            \\
    \  &     225     &       1.1       & \     &      4.8             & \                        & \                  \\
    \  &     265     &       0.93      & \     &      7.8             & \                        & \                  \\ \hline
  APEX &     150     &       0.8       & 2006  & \multicolumn{3}{c|}{}\\
    \  &     217     &       0.8       & \     & \multicolumn{3}{c|}{}\\ \hline
   SPT &     150     &       1         &2007   &      10              & 11                       &      \             \\ 
    \  &     220     &       0.7       & \     &      60              & \                        &     4000           \\
    \  &     275     &       0.6       & \     &      100             & \                        &       \            \\ \hline
   Planck &  143     &       7.1       & 2008  &       6              & 0.35                     &       \            \\
     \    &  217     &       5         & \     &       13             & \                        &     40000          \\
     \    &  353     &       5         & \     &       40             & \                        &       \            \\ \hline
\end{tabular}
\caption[Experiments]
{Characteristics of some of the large SZ surveying instruments.  We give the instrumental noise, counts detected at $S/N>5$ with our 
matched filter and survey area for the 4 experiments that we have simulated in detail, namely AMI, ACT, SPT \& Planck (see text).  
A list of web pages for these and other experiments is given in the reference section.
Adapted from Melin et al. (2006)~\cite{melin_etal}.
{\bf Note added:} The new WMAP3 results were published at the time of writing.  The lower value of 
$\sigma_8$ favored by the new release can lower the predicted cluster counts by up to a factor $\sim 2$. 
We are evaluating the changes for each
experiment in detail.
}
\label{tab:experiments}
\end{table}

\begin{figure}
\includegraphics[scale=0.45]{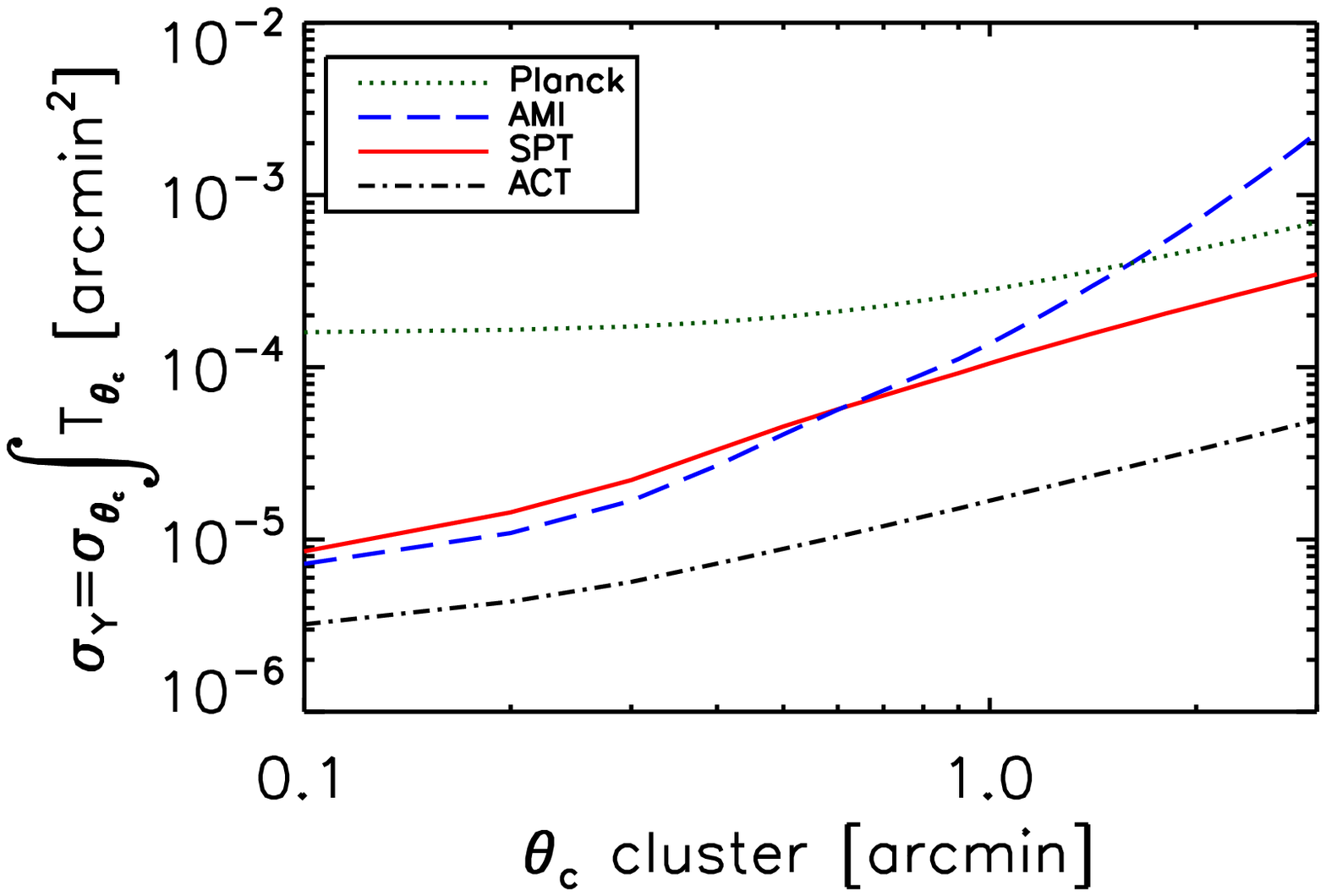}
\includegraphics[scale=0.45]{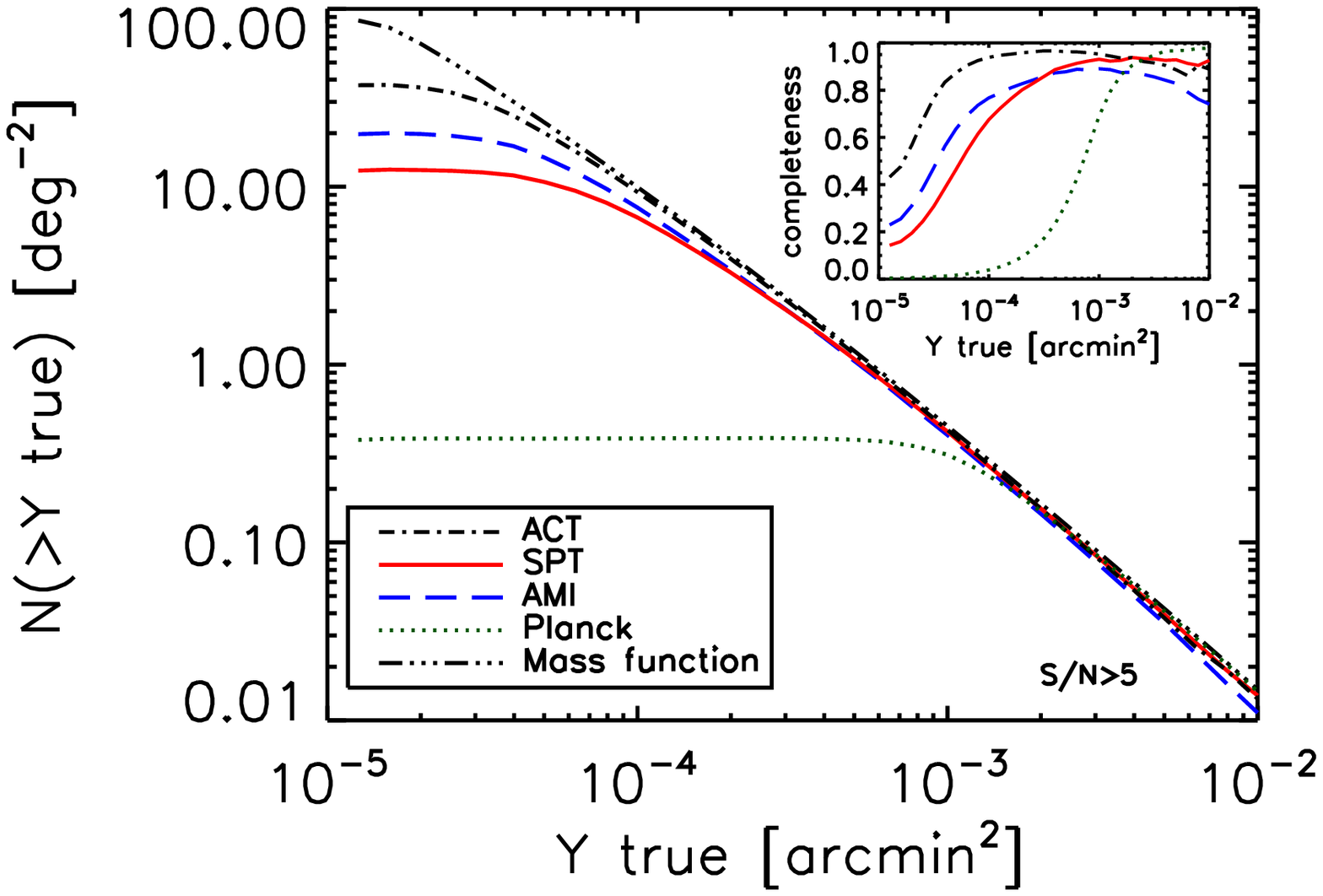}
\caption{The {\em left--hand panel} plots the matched filter noise $\sigY$ 
as a function of filter scale $\thetac$ (core radius of a cluster matched to the filter) 
for different surveys, as labeled.  The filter noise is generated by primary CMB 
anisotropy and instrumental noise.  Clusters lying above the curve of a particular
experiment have $S/N > 1$.  Integrated source counts at $S/N>5$ for each survey are shown in the 
{\em right--hand panel}, along with the simulation input counts (curve labeled ``mass function'').
We give the catalog completeness percentage (ratio of the experimental curve to the input mass 
function counts) in the inset.  The important point is that the surveys are not flux limited, and
are significantly incomplete even at 5 times their point source sensitivities (5 times the y--intercept 
of the left--hand curves).  Adapted from Melin et al. (2006). 
{\bf Note added:} The new WMAP3 results were published at the time of writing.  The lower value of 
$\sigma_8$ favored by the new release can lower the predicted cluster counts by up to a factor $\sim 2$.  
We are evaluating the changes for each
experiment in detail.
}
\label{fig:completeness}
\end{figure}

\begin{figure}
\includegraphics[scale=0.45]{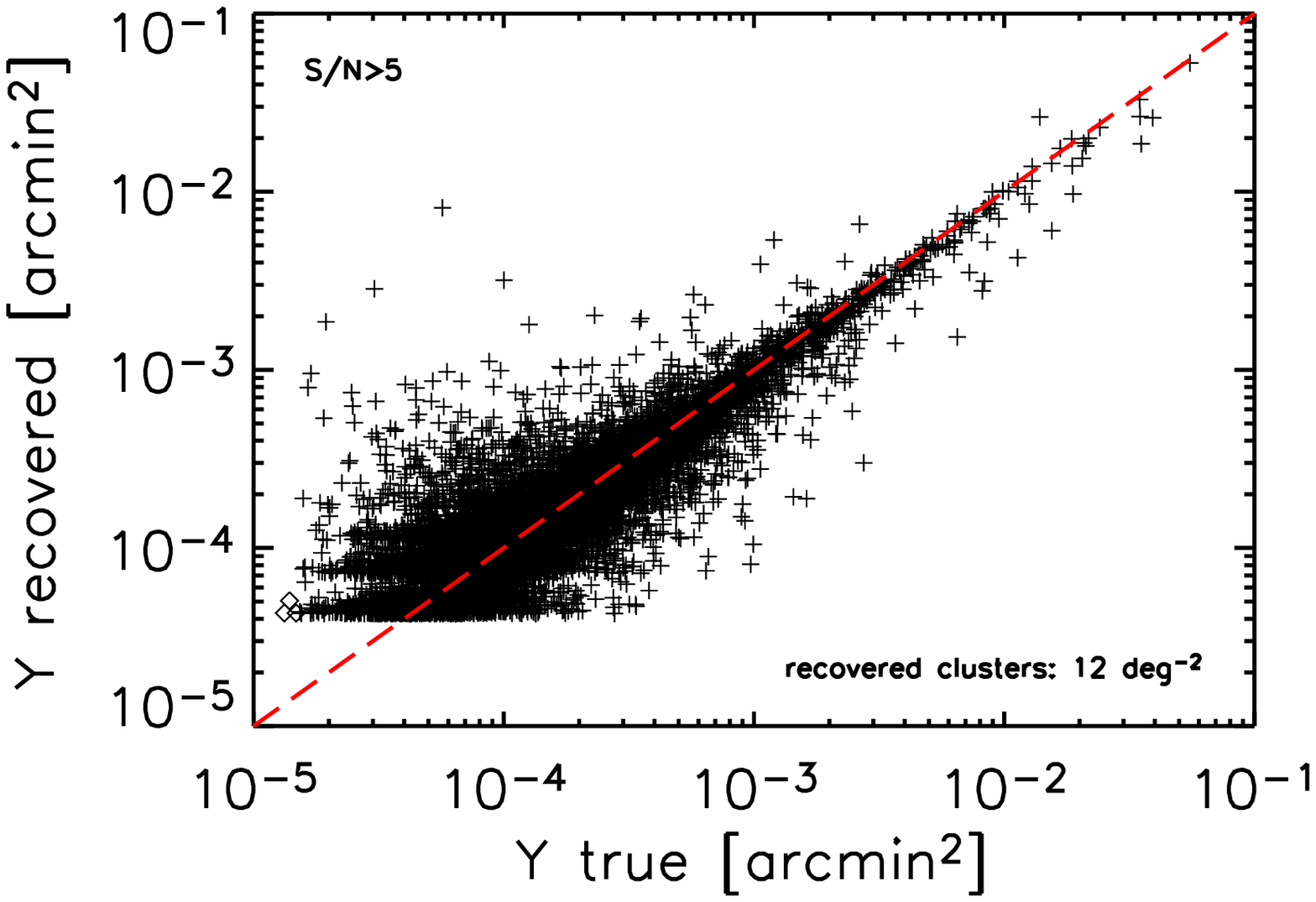}
\includegraphics[scale=0.45]{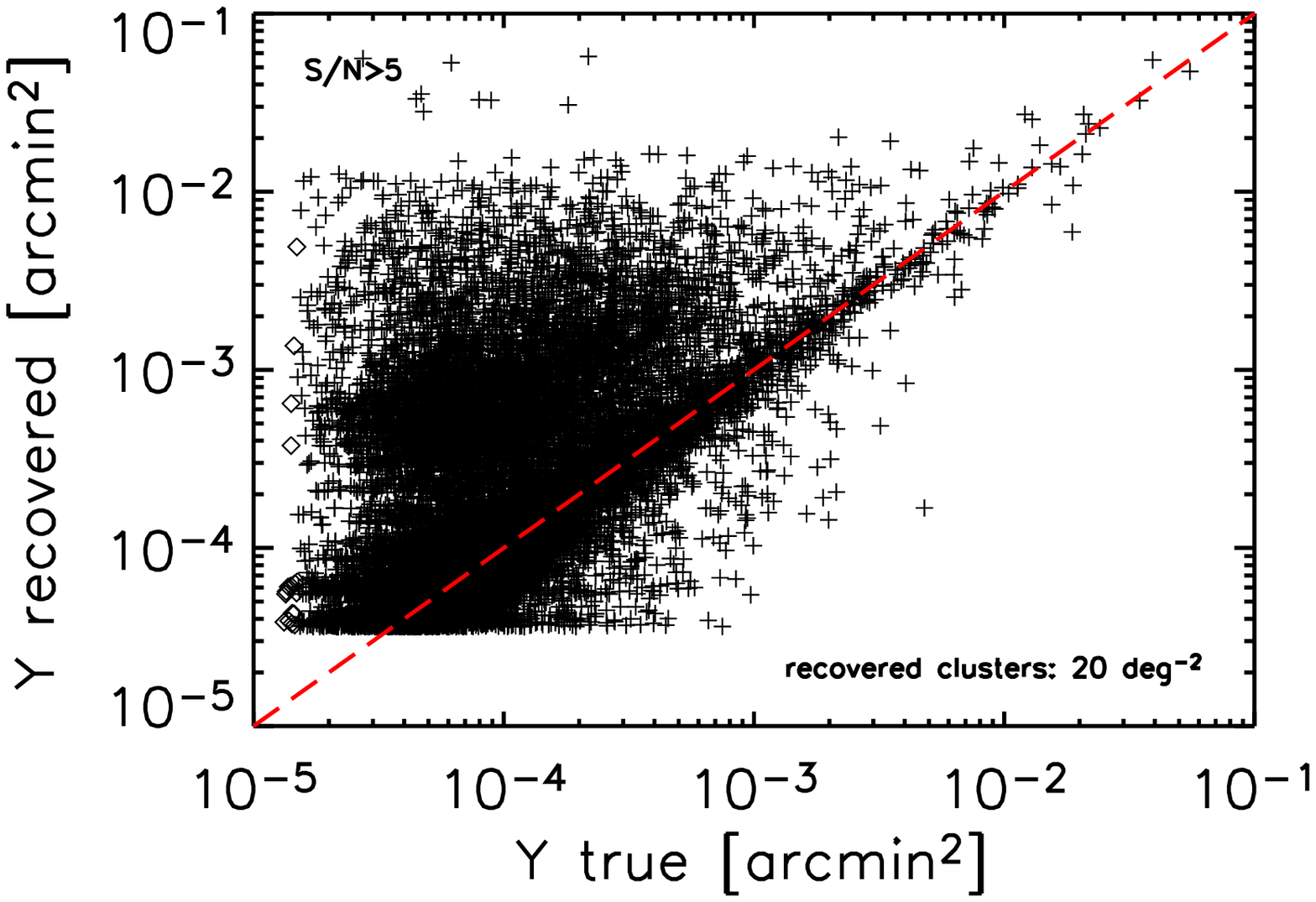}
\caption{Photometric recovery in terms of integrated Compton $Y$ parameter for SPT (left) and
AMI (right).  Compton $Y$ values (in arcmin$^2$) recovered by the matched filter are plotted against the
input $Y$ values taken from the simulation catalog.  Each point represents a single cluster detected at $S/N>5$.  
The red dashed curve gives the equality line.  For SPT the characteristic scatter at fixed $Y_{\rm true}$ is 
$\sim 40$\%.  Confusion with primary CMB anisotropy seriously compromises photometric recovery 
of the single frequency survey (chosen here as AMI). Adapted from Melin et al. (2006).
}
\label{fig:photometry}
\end{figure}

\section{SZ Catalog Construction}
In anticipation of the upcoming surveys, we have undertaken an extensive simulation effort to study
survey selection functions and observational uncertainties, and I will report here two key
results from this work to date.  Details can be found in Melin {\it et al.}~\cite{melin_etal} 
and Bartlett \& Melin~\cite{bart_melin}.

We have developed a cluster detection algorithm based
on a filter matched to both the spectral ($\jnu$) and spatial (isothermal $\beta$--model) signature of the cluster 
SZ signal~\cite{matched_filters}.  
The filter is designed to rapidly and optimally extract clusters of all sizes from the 
astrophysical foregrounds.  Rapidity is essential because we want to simulate a given survey 
many times in order to accurately quantify selection effects and observational uncertainties.

Applying the detection routine to Monte Carlo simulations of the AMI, ACT, Planck and SPT surveys, we calculated the 
survey sensitivities, counts and completeness values shown in Figure~\ref{fig:completeness}.  In this Figure, $\sigY$ is 
the total noise through the matched filter generated by both instrumental noise and primary CMB anisotropy (for the standard 
WMAP cosmology), and $\thetac$ refers to the core radius of our simulated SZ 
profiles truncated at 10 core radii.  The curves correspond to the different surveys as labeled.  A cluster 
observed by SPT, for example, with angular core radius $\thetac$ and $Y=\sigY(\thetac)$, i.e., on the solid, red
line, would have a $S/N=1$.  

We see from this Figure that the surveys are not simply limited in 
flux; cluster selection depends on both flux and angular size.  This is particularly true for the 
high resolution ground--based surveys, while the lower resolution of Planck yields a nearly
flux--limited catalog.  In this Figure we only show the effects of instrumental noise and CMB
anisotropies.  In Bartlett \& Melin~\cite{bart_melin} we examine the additional effects caused by 
extragalactic point sources.

In the right--hand panel I show the integrated counts of clusters detected at $S/N>5$.  These
are objects to be found in the left--hand panel above the corresponding curves, each displaced upward by
a factor of 5.  The counts roll away from the input counts (labeled {\em mass function}) at the faint end
as the catalogs become incomplete.  In the inset, we see the completeness defined as the ratio of 
detected to input clusters (ratio of one of the experimental curves to the mass function curve).
It is important to note that all of the surveys are significantly incomplete even at 5 times their 
point--source sensitivities, which are given by 5 times the y--intercept of the curves shown in the
left--hand panel.  This just illustrates again that the surveys are not simply flux limited.
 
I now turn to another important, but often neglected issue -- photometric recovery.  In Figure~\ref{fig:photometry} I plot
recovered $Y$ versus input $Y$ for clusters detected at $S/N>5$ in SPT (left--hand panel) and AMI (right--hand panel);
the Planck case is similar to SPT.  We see that when applied to SPT our algorithm gives an unbiased estimate of 
the true SZ flux, but with $\sim 40$\% scatter (at fixed $Y_{\rm true}$), which is much larger than the 
intrinsic scatter predicted by numerical simulations~\cite{motl_etal05}.  
The photometry of the single frequency AMI survey, on the other hand, is very seriously compromised; the distribution
at fixed $Y_{\rm true}$ is in fact bimodal.  We have traced this effect to confusion with primary CMB 
anisotropy that is unavoidable with single frequency observations.  The conclusion is that such single
frequency surveys must follow--up their clusters at a second frequency in order to eliminate the confusion.

\section{Conclusions}

Future galaxy cluster surveys will provide a wealth of information on dark energy, dark matter 
and structure and galaxy formation.  Among these surveys, those based on the SZ signal will profit
from its intrinsic ability to find clusters at high redshift and its expected tight correlation 
to halo mass.  Over the next 5 years, these surveys will provide large, well--defined (in terms of mass) 
catalogs containing hundreds to thousands of clusters at redshifts beyond unity or, in other words, 
multiplying by 10--100 the number of known clusters at these redshifts\footnote{The new WMAP3 
results were published at the time of writing.  The lower value of 
$\sigma_8$ favored by the new release can lower the predicted cluster counts by up to a factor $\sim 2$.  
We are evaluating the changes for each
experiment in detail.}.  This 
will give us a new view of galaxy formation in dense environments and a measure of the expansion 
rate at this crucial epoch marking the transition between matter and dark energy domination.  

Many of the surveys will begin this year (2006), and a number of important surveying issues
require further study.  These include accurate evaluation of survey selection functions and
observational errors.  I have shown two results from our studies based on a matched filter
detection algorithm and simulations of different surveys.  We find that high resolution ground--based
surveys select clusters not simply on flux, but on a combination of flux and angular size, and that 
this must be properly accounted for when modeling cluster counts.  We also find that observational 
scatter on measured cluster flux $Y$ (e.g., $\sim 40\%$ in the case of SPT) largely exceeds the intrinsic 
scatter predicted by numerical simulations~\cite{motl_etal05}.  Furthermore, confusion with
primary CMB anisotropy severely compromises photometry in single frequency observations; follow--up 
at a second frequency (or in X--rays) will therefore be necessary for these surveys.

\section*{Acknowledgments}
I wish to thank the organizers of the {\em Rencontres de Moriond} for a pleasant meeting, and P. Richards for 
a very instructive discussion on bolometer surveying sensitivity.  I also thank my collaborators J.--B. Melin and J. Delabrouille.

\end{document}